\documentclass[a4paper,showpacs]{revtex4}
\usepackage{amsfonts,amsmath,epsfig}
\usepackage{hyperref}
\newcommand{\be}{\begin{equation}}
\newcommand{\ee}{\end{equation}}
\begin{document}
\title{Electron wave functions in a magnetic field}
\author{D.~K.~Sunko}
\email{dks@phy.hr}
\affiliation{Department of Physics, Faculty of Science, University of
Zagreb,\\ Bijeni\v cka cesta 32, HR-10000 Zagreb, Croatia.}

\begin{abstract} The problem of a single electron in a magnetic field is
revisited from first principles. It is shown that the standard quantization,
used by Landau, is inconsistent for this problem, whence Landau's wave
functions spontaneously break the gauge symmetry of translations in the plane.
Because of this Landau's (and Fock's) wave functions have a spurious second
quantum number. The one-body wave function of the physical orbit, with only
one quantum number, is derived, and expressed as a superposition of Landau's
wave functions. Conversely, it is shown that Landau's wave functions are a
limiting case of physical solutions of a different problem, where two quantum
numbers naturally appear. When the translation gauge symmetry is respected,
the degeneracy related to the choice of orbit center does not appear in the
one-body problem. \end{abstract}


\maketitle

\section{Introduction}

The motion of a free electron in a homogeneous magnetic field is a classic
problem in elementary quantum mechanics. It was first treated by
Landau~\cite{Landau30} in 1930, and remains a subject of active research to
this day~\cite{Govaerts09}. The wave functions derived by Landau are routinely
used as building blocks of many-body states in a magnetic
field~\cite{Abergel09}.

Canonical quantization, as established by Dirac~\cite{Dirac30}, is the only
impeccable way to quantize a classical system, against which all other
approaches are measured. The method consists in promoting Poisson brackets of
classical mechanics into commutators by the prescription
$\{x,p_x\}=1\rightarrow [\hat{x},\hat{p}_x]=i\hbar$, where I have taken the
standard example of position and momentum dynamical variables. This step
establishes the structure of the quantum theory, but leaves the freedom to
concretely realize the Hilbert space on which the operators $\hat{x}$ and
$\hat{p}_x$ act. The standard choice is $\hat{x}=x$, a real number, and
$\hat{p}_x=-i\hbar\partial_x$, turning them into operators on the Hilbert
space $\mathbf{L}^2(\mathbb{R})$ of square-integrable functions on the real
line. The canonical quantization, when taken together with this particular
realization of Hilbert space, is usually called ``first quantization," and the
corresponding wave functions are ``Schr\"odinger wave functions." Before the
work of Dirac, quantization was simply understood to be the replacement
$p_x\rightarrow -i\hbar\partial_x$ in the classical Hamiltonian, irrespective
of the Poisson bracket structure, a prescription called ``naive quantization."

Canonical and naive quantization should in principle give the same result,
because whichever canonical transformation can be made at the level of Poisson
brackets, can also be made at the level of commutators, after quantization.
Exceptionally, when there is a hidden gauge degree of freedom at the classical
level, naive quantization overcounts the physical degrees of freedom in the
system. The purpose of the present paper is to show that this is the case with
a free electron in a magnetic field, with translations as the relevant gauge
group. The situation is closely analogous to the ground state of the hydrogen
atom, which may similarly be considered to have a huge degeneracy, coming from
the different positions of the proton in space, but which is easily separated
in the calculation from the one-body electron states within the atom. For an
electron in a magnetic field, the separation is more difficult. The physical
reason for the difficulty is that, unlike the atomic electron, the free
electron chooses its own orbit-center, so the calculation has to be set up in
a way invariant to that choice, in the absence of adiabatic factorization. The
formal reason is that naive quantization is inconsistent in the presence of a
gauge degree of freedom at the classical level, as first noticed by
Dirac~\cite{Dirac50}. Naive quantization subtly introduces the positional
degeneracy of the orbit-center into the calculation, which appears as a
spurious constant of motion (wave-number) in the one-body wave function. An
explicit integration over the spurious wave-number, Eq.~\eqref{lansol} below,
is then needed to restore the translation-gauge symmetry and give the true
one-body wave function, which is non-degenerate. Canonical quantization
clearly separates the one-body and many-body aspects of the problem: there is
no degeneracy in the former, while in the latter, it appears manifestly as
translation-gauge fixing, associated with choosing the position of the orbit
center explicitly. It is shown below that the well-established coherent-state
formalism~\cite{Feldman70} is nothing but the most natural implementation of
such translation-gauge fixing.

The Hilbert space established by first quantization is not the only one
possible. Any particular choice is justified by canonical quantization if it
is formally realized as a transform of the first-quantized Hilbert space. An
important example, needed in the following, is the Fock space
$\mathbf{F}(\mathbb{C})$, spanned by entire analytic functions $f(z)$,
$z=x+iy\in\mathbb{C}$, with the property $\int dxdy\,|f(z)|^2e^{-|z|^2} <
\infty$. It is obtained from 1D first-quantized space by the Bargmann
transform~\cite{Bargmann61},
\be
f(z)=\frac{1}{\pi^{3/4}}\int_{-\infty}^{+\infty}
\exp\left(-\frac{z^2+x^2}{2}+xz\sqrt{2}\right)\Psi(x)dx,
\label{bargtrans}
\ee
where $\Psi(x)\in\mathbf{L}^2(\mathbb{R})$ is a Schr\"odinger wave function.
In particular, if $\Psi(x)$ is the Hermite function $\psi_n(x)$, the
well-known Schr\"odinger eigenfunction of the oscillator, then $f(z)\sim z^n$,
while the oscillator Hamiltonian becomes
\be
\left(z\partial_z+\frac{1}{2}\right)\hbar\omega
\label{bargham}
\ee
under the same transform. Clearly, the entire functions $z^n$ are its
eigenfunctions. Notably, if the real variable $x$ in the integral
transform~\eqref{bargtrans} is a point in real space, then $z=x+iy$ cannot be
interpreted to refer to real space.

\section{The Fock problem}

In order to establish the paradigm of canonical quantization, needed to solve
the Landau problem~\cite{Landau30}, I invoke the related Fock
problem~\cite{Fock28}, of a particle of charge $q=-e<0$ in a 2D isotropic
harmonic oscillator potential, subject to a perpendicular magnetic field. The
classical Hamiltonian of Fock's problem may be written
\be
H_F=\frac{\omega_+}{2}\left(v_x^2+v_y^2\right)+
\frac{\omega_-}{2}\left(x_0^2+y_0^2\right),
\label{classfock}
\ee
where the symmetric gauge is assumed throughout, and
\be
\begin{pmatrix}
x_0\\y_0\\v_x\\v_y
\end{pmatrix}=
\begin{pmatrix}
\alpha/2&0&0&-1/\alpha\\
0&\alpha/2&1/\alpha&0\\
0&-\alpha/2&1/\alpha&0\\
\alpha/2&0&0&1/\alpha
\end{pmatrix}
\begin{pmatrix}
x\\y\\p_x\\p_y
\end{pmatrix},
\label{carttrans}
\ee
with $\alpha=\sqrt{M\Omega}$, is a canonical transformation of the original
dynamical variables. Here $\{x_0,y_0\}=\{v_y,v_x\}=1$, and all other Poisson
brackets are zero. The frequencies are
$
\omega_\pm=(\sqrt{\omega_c^2+4\omega_0^2}\pm \omega_c)/2
\equiv (\Omega\pm\omega_c)/2
$,
where $\omega_0$ is the intrinsic frequency of the oscillator, and
$\omega_c=eB_0/Mc$ is the cyclotron frequency of the particle.

The classical Fock problem is easy to solve. The trajectory is epicyclic. It
consists of a ``fast" counterclockwise rotation, with angular frequency
$\omega_+$, on a circle, combined with a ``slow" clockwise rotation of the
circle center, with frequency $\omega_-$, around the origin of the harmonic
potential. To quantize it, first rescale $v_y$ to $v$, given by
\be
\frac{v_y}{\sqrt{\hbar}}
\longrightarrow
\frac{x+k_y}{\sqrt{2}}
\equiv v,
\ee
where the arrow means rescaling by $r_0=l_0\sqrt{2}=\sqrt{2\hbar c/eB_0}$, in
particular $k_y=r_0p_y/\hbar$. The conjugate variable in the Poisson bracket
is now replaced by canonical prescription, $v_x/\sqrt{\hbar}\rightarrow
-i\partial_v$, so the first-quantized Hamiltonian is
\be
\widehat{H}_F=\frac{\hbar\omega_+}{2}
\left(-\partial_{vv}
+v^2\right)+
\frac{\hbar\omega_-}{2}
\left(-\partial_{ww}+w^2\right),
\label{fockham}
\ee
where $w=(x-k_y)/\sqrt{2}$ is the similarly rescaled $x_0$. Evidently its
solution (the Schr\"odinger wave function) is
\be
\psi_{n_+}(v)\psi_{n_-}(w),
\label{focksch}
\ee
where the $\psi_n$ are Hermite functions. The spectrum is
$
E=(n_++1/2)\hbar\omega_++(n_-+1/2)\hbar\omega_-
$,
as obtained by Fock. However, Fock used naive quantization, so that his wave
function was directly obtained in real space. Here the canonical quantization
has imposed a mixed first-quantized space, real space in one direction and
momentum space in the other. In order to compare it with Fock's solution, one
needs to transform this solution into real space, by a Fourier transform:
\be
\int dk_ye^{iyk_y}\psi_{n_+}(\frac{x+k_y}{\sqrt{2}})
\psi_{n_-}(\frac{x-k_y}{\sqrt{2}})\sim
(x+iy)^{n_+-n_-}L_{n_-}^{n_+-n_-}(x^2+y^2)
e^{-(x^2+y^2)/2},
\label{focksol}
\ee
which is precisely the wave function expressed in Laguerre functions,
found by Fock~\cite{Fock28}. The mathematical interpretation of Laguerre
functions as Wigner transforms of Hermite functions is well known in the
representation theory of the Heisenberg group~\cite{Thangavelu93}. Here a
physical interpretation is obtained, that they are Fourier transforms of the
Schr\"odinger wave functions, Eq.~\eqref{focksch}, which describe the two
connected rotations of Fock's problem in first-quantized space.

For Fock's problem, naive quantization is justified by the canonical
prescription. The advantage of choosing the first-quantized space in accord
with the Poisson-bracket structure of the classical problem is better physical
insight. In Fock's case, the natural independent variables are the velocity
component $v_y$ and orbit center component $x_0$, both of which oscillate
harmonically, because they are 1D projections of the corresponding circular
motions. This is clear from the Schr\"odinger wave functions~\eqref{focksch},
which are elements of $\mathbf{L}^2(\mathbb{R}^2)$ in the plane $(v,w)$. The
wave function space $\mathbf{L}^2(\mathbb{R}^2)$ in the physical plane $(x,y)$
is an alternative choice of Hilbert space. The Hamiltonian, obtained by
Fourier-transforming Eq.~\eqref{fockham}, is the same as obtained from the
classical Hamiltonian by naive quantization, but, as we shall see now, that is
not by itself a justification of naive quantization.

\section{The Landau problem}

Having established the quantization method, and checked it on Fock's problem,
I now apply it to Landau's problem~\cite{Landau30} of an electron in empty
space, subject to a constant magnetic field $\mathbf{B}=B_0\hat{\mathbf{z}}$.
Leaving implicit the motion along the field, the quantization in the
perpendicular plane can be obtained by setting $\omega_0=0$ in Fock's problem,
whence $\Omega=\omega_+=\omega_c$, and $\omega_-=0$. The first-quantized
Hamiltonian becomes
\be
\widehat{H}_L=\frac{\hbar\omega_c}{2}
\left(-\partial_{vv}+v^2\right),
\label{lanham}
\ee
because the orbit center has stopped moving. Significantly, the velocity
component $v_y$, which appears here, has no reference to any particular
position in space. The Schr\"odinger wave function is $\psi_n(v)$, which
refers only to the ``internal" motion of the electron, i.e.\ relative to a
given center, wherever it may be. It is obviously non-degenerate, and means
that the projection $v_y$ of the velocity on the $y$-axis oscillates
harmonically, which is, of course, true for uniform circular motion. The
corresponding real-space wave function is obtained by the same Fourier
transform, from velocity space to real space, as in Fock's problem, which
gives (scaling by $r_0=l_0\sqrt{2}$)
\be
\int dk_ye^{iyk_y}\psi_{n}(\frac{x+k_y}{\sqrt{2}})
\sim e^{-ixy}\psi_n(y\sqrt{2}),
\label{lansol}
\ee
because Hermite functions are eigenfunctions of the Fourier
transform~\cite{Thangavelu93}. This is the correct physical solution of the
one-body Landau problem, and the central result of the present paper. In the
Landau gauge and dimensionful variables, it is just $\psi_n(y/l_0)$. Note that
Landau's original wave function, $e^{iyk_y}\psi_{n}((x+k_y)/\sqrt{2})$,
reappears as its Fourier component, so the physical solution is a
superposition of Landau's formal solutions as basis vectors: the latter are
``harmonics'' of the physical orbit. This means that Landau's wave function is
not in real space: its variables are $x$ and $k_y$, while $y$ is a parameter.
It only appears to be in real space ($x$ and $y$ variables, $k_y$ a parameter)
if one tries to solve the naively quantized Hamiltonian by direct attack,
because then one is working ``blindly" in the whole space
$\mathbf{L}^2(\mathbb{R}^2)$ of wave functions in the plane, while the
Schr\"odinger wave functions of Eq.~\eqref{lanham} are in
$\mathbf{L}^2(\mathbb{R})$, because the Landau Hamiltonian has only one degree
of freedom. In particular, it is misleading to relabel $k_y$ as $-x'$ (say),
which is commonly done in the literature, without taking into account that it
is also a wave-number label. It implies that, although Landau's wave functions
are a basis in real space, a physical wave function of a single electron
cannot be described by a single Landau vector, because for an electron
orbiting around a stationary center, the linear wave-number is not a good
quantum number. Eq.~\eqref{lansol} is precisely the summation over this
spurious constant of motion, required to produce a physical solution, whose
orbit center is fixed. The dual role of $k_y$ will be fully explained in
the next section.

Eq.~\eqref{lansol} is not symmetric with respect to the Cartesian axes. This
is a reflection of the ``handedness'' of the original problem, formally
encoded in the signs of the Poisson brackets: $\{v_y,v_x\}=1$, rather than
$-1$. Indeed, for a positively charged particle the $x$- and $y$-axes
exchange place. In effect, the time-reversal symmetry breaking of the magnetic
field is reduced to the statement that the Hamiltonian~\eqref{lanham}
describes the oscillation of the $y$-component of the velocity, rather than
the $x$-component. It is particularly gratifying in this context that the
Landau gauge-fixing term $e^{-ixy}$ appears as a result in Eq.~\eqref{lansol},
namely as the eigenvalue of the Fourier transform, rather than an assumption.
Hence the preference of one axis over the other is not due to any a priori
choice to search for the solution in the Landau gauge. Because
$v_y=(x+k_y)/\sqrt{2}$, one can absorb the gauge-fixing term into the
velocity, so the same Eq.~\eqref{lansol} can be written
\be
\psi_n(y\sqrt{2})\sim\int dv_ye^{iv_yy\sqrt{2}}\psi_{n}(v_y),
\label{langsol}
\ee
showing that the Landau-gauge solution in real space is nothing but a Fourier
transform directly from velocity space. Physically, this equation means that
harmonic oscillation in the $y$-component of the velocity is equivalent to
harmonic oscillation in the $y$-coordinate of the position. The gauge-fixing
term appears as an independent factor only if one insists on writing the
Fourier transform in the wave number $k_y$, in spite of the velocity being the
canonical variable of the problem: $iy\cdot v_y\sim iy\cdot(x+k_y)\sim iy\cdot
x+iy\cdot k_y$. A Landau ``solution'' (basis vector) in the Landau
gauge is just a single term under the integral in Eq.~\eqref{langsol}, as can
be seen by rewriting it in terms of $k_y$ instead.

Notably, if the Hamiltonian~\eqref{lanham} is Fourier-transformed to real
space, one does obtain, as in Fock's case, the same Hamiltonian as by naive
quantization. However, its solutions are limited by the same transform to the
highly restricted class of Eq.~\eqref{lansol}, meaning that the transform to
real space is also a prescription to embed $\mathbf{L}^2(\mathbb{R})$ into
$\mathbf{L}^2(\mathbb{R}^2)$. Physically, the 1D oscillation of the projection
of a circular orbit is being expressed in the larger wave-function space,
capable of carrying true 2D oscillation.

In Landau's approach, $n$ and $k_y$ are quantum numbers, so the above
canonical approach removes the infinite degeneracy in the second quantum
number. This is appropriate, because the two degrees of freedom in Fock's
problem are reduced to one in Landau's problem. It may similarly be noted that
when $\omega_-=0$, the solution to Fock's problem, Eq.~\eqref{focksol},
becomes infinitely degenerate in the quantum number $n_-$, and may also serve
as a formal solution (basis) for Landau's problem. However, it cannot be the
physical solution, formally because a problem with one degree of freedom
cannot have two quantum numbers, and physically because it describes a circle
whose center also moves on a circle, while in the Landau problem the center is
motionless.

The last observation points to a way to guess the subclass of solutions to
Fock's problem which are also solutions of Landau's problem: simply put
$n_-=0$ in Eq.~\eqref{focksol}. This immediately gives ($n_+=n$)
\be
(x+iy)^ne^{-(x^2+y^2)/2},
\label{z00}
\ee
which agrees with the Fock-space solution of a single harmonic oscillator,
given in the Introduction. This solution is not in Fock space, but still in
real space, as indicated by the Gaussian factor (absent in Fock space).
Alternatively, the Bargmann transform of the Hamiltonian~\eqref{lanham} from
the variable $v$ to the variable $z$ gives the Fock-space solution $z^n$
directly from $\psi_n(v)$. This justifies the guess~\eqref{z00}, showing also
that in this case the Fock-space $z=x+iy$ does refer to a position in real
space, which is possible because the original variable $v$ was not itself in
real space.

Both the real-space solution~\eqref{lansol} and the Fock-space $(x+iy)^n$ are
just different images, under the Fourier and Bargmann transforms respectively,
of the original solution $\psi_n(v)$ of the one-particle
Hamiltonian~\eqref{lanham} in velocity space, so they naturally inherit its
non-degeneracy. To put these one-particle results in the many-body context,
and make contact with well-known results, note that there is still expected a
\emph{parametric} infinite degeneracy of the problem in real space, because
the orbit center can be anywhere in the plane. This degeneracy is physically
different than degeneracy in a quantum number, and in particular the parameter
denoting the orbit center must be a continuous $c$-number, with a sharp value.
To identify it in the real-space representation, write an \emph{ansatz} for
the wave function as
\be
\psi(z,\bar{z})=u(z,\bar{z})e^{-|z|^2/2}
\label{zzans}
\ee
where $z,\bar{z}=x\pm iy$ is just a coordinate transformation here, so $x$ and
$y$ do refer to a real-space position. The real-space eigenvalue equation
for $u$ becomes
\be
-u_{z\bar{z}}+zu_z=nu,
\label{hzz}
\ee
where $n=E/(\hbar\omega_c)-1/2$. Solving it by separation of variables,
$u(z,\bar{z})=f(z)g(\bar{z})$, one obtains quite generally
\be
u(z,\bar{z})=(z-z_0)^ne^{z_0\bar{z}},
\label{gtrans}
\ee
where $z_0$ is the separation constant, a complex number. Notably, this
separation of variables yields a physically different set of solutions than
the usual $u(z,\bar{z})=f(|z|)g(\arg z)$, which leads to Fock's
solution~\eqref{focksol}. Eq.~\eqref{gtrans} is a particular resummation of an
infinite series in the `moving-center' solutions~\eqref{focksol} over the
quantum number $n_-$, which produces an electron orbit with a
center fixed at $z_0$. Such a construction is analogous to the
expression~\eqref{lansol}. When $z_0=0$, the series collapses to a single
term, Fock's solution~\eqref{focksol} with $n_-=0$, as guessed above. These
eigenstates are also coherent states, and were used for that reason, in the
context of many-body path integrals~\cite{Girvin84,Kivelson86}. As coherent
states, they are complete precisely when the density of the orbit-center grid
$z_0$ is one orbit per flux quantum~\cite{Feldman70}, which is an independent
confirmation of Landau's estimate of the degeneracy of Landau levels. This
discretization of the separation constant $z_0$ plays the role of a spectrum.

Parenthetically, it is unexpected for coherent states to be simultaneously the
eigenstates of the Hamiltonian. Usually, they are only eigenstates of some
lowering operator. The states~\eqref{gtrans} achieve this surprising duality
by being eigenstates of the first term in Eq.~\eqref{fockham}, and coherent in
the second term, which is zero in the Landau Hamiltonian. The corresponding
lowering operator is $\hat{x}_0+i\hat{y}_0$, physically the orbit-center
position, with eigenvalue $z_0$.

The term $e^{z_0\bar{z}}$ appears above as the extra factor in the wave
function, needed to compensate the action of translating the physical orbit in
the plane, $z\rightarrow z-z_0$. In other words, it is a gauge-fixing term,
with translations as the underlying gauge group. The shift of the orbit center
by $z_0$ comes after the choice of origin of coordinates has been fixed by the
zero of the vector potential, which is where the Gaussian envelope in
Eqs.~\eqref{lansol} and~\eqref{zzans} is centered. Shifting the origin of
coordinates shifts the potential by a constant, which is an electromagnetic
gauge transformation. Thus translation and electromagnetic gauge fixing appear
independently in the wave function. In gauge-theory idiom, motion in $z$ is
physical, while motion in $\bar{z}$ is ``pure gauge." It is possible to write
the same states in the form of shifted Gaussians with an extra phase
factor~\cite{Kivelson86}, which however hides the natural structure of the two
distinct gauge groups. The separation of physical and gauge motion plays a
striking role in the demonstration that the Landau-level quantum number $n$ is
just the physical angular momentum around the arbitrary orbit center, as shown
in the Appendix. Because of that interpretation, the preference of $z$ over
$\bar{z}$ in the holomorphic coordinates is physically more intuitive than the
preference of $y$ over $x$ in the Cartesian coordinates, although both are
equally due to the time-reversal symmetry breaking. In this context
Eq.~\eqref{lansol} is a ``Cartesian'' coherent state, to be compared with the
``holomorphic'' one in Eq.~\eqref{gtrans}. The reduction to one coordinate
reflects the loss of the orbit-center degree of freedom in both cases.

The translation gauge reappeared in Eq.~\eqref{gtrans} only because the wave
equation in real space was being solved by direct attack. In principle, one
should not do that with a transformed one-body equation. Instead, all of its
physically distinct solutions are to be generated from the first-quantized
solutions by the self-same transform. Technically, this correctly embeds the
space $\mathbf{L}^2(\mathbb{R})$, of Schr\"odinger wave functions which solve
the Landau problem, into the space $\mathbf{L}^2(\mathbb{R}^2)$, of all
possible solutions of the two-dimensional real-space Hamiltonian. Physically,
the resummation in Eq.~\eqref{lansol} may be considered as restoration of the
broken translation-gauge symmetry.

The translation-gauge degree of freedom becomes physical, however, when one
considers the many-body problem. The localized electrons then fill the
translation-gauge spectrum, given by the discretized orbit-center grid $z_0$.
The energy spectrum, on the other hand, is given by the first-quantized
Hamiltonian~\eqref{lanham}, free of the translation gauge.

\section{A problem solved by Landau-like wave functions}

As shown in Eq.~\eqref{lansol} above, Landau's wave functions are
eigenfunctions of the Landau problem, but not, individually, solutions.
Similarly, Fock's wave functions can be a basis for Landau's problem, but they
only solve Fock's problem: a single electron can be in a single Fock basis
state only if there is an isotropic 2D harmonic potential present, otherwise
the circular motion of the orbit center is unphysical. In this section, it
will be formally shown that the wave quantum number in Landau's wave functions
is inherited from a related problem, where it refers to a physical linear
charge current coming from orbit center motion, just like the angular momentum
quantum number $n_-$ in Fock's problem refers to a clockwise circular current.

\begin{figure}
\includegraphics[height=3cm]{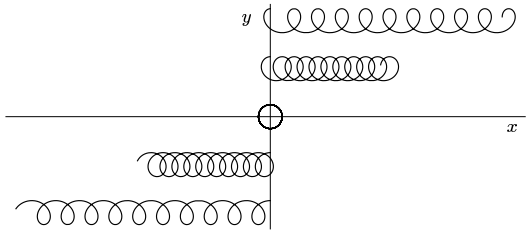}
\caption{Several trajectories of the class \eqref{sol1d}, evolved for the same
length of time. Here $\omega_c/\omega_0=5$, or $\omega_c/\omega\approx 0.98$.}
\label{figlandau}
\end{figure}
The intuitive analogy with Fock's problem suggests to add a 1D harmonic
potential to the magnetic field in Landau's problem. Orienting the oscillator
minimum along the $x$-axis, the classical Hamiltonian is
\begin{align}
H&=\frac{1}{2}M(\dot{x}^2+\dot{y}^2)+\frac{M\omega_0^2}{2}y^2\nonumber
\\
&=\frac{1}{2M}(p_x^2+p_y^2)+\frac{\omega_c}{2}(xp_y-yp_x)+
\frac{M\omega_c^2}{8}x^2+\frac{M\Omega^2}{8}y^2,
\label{ham1d}
\end{align}
where $\Omega^2=\omega_c^2+4\omega_0^2$. A representative class of
trajectories starts on the $y$-axis with velocity horizontal; they are
\be
x(t)=V_xt-R\frac{\omega_c}{\omega}\sin\omega t,\qquad
y(t)=Y_0+R\cos\omega t,
\label{sol1d}
\ee
where $\omega^2=\omega_c^2+\omega_0^2$, and, importantly,
\be
V_x=\frac{\omega_0^2}{\omega_c}Y_0.
\label{constr}
\ee
This constraint already indicates the dual role played by the wave-number in
Landau's wave function, where, as shown in the following, it increases with
the distance from the oscillator axis to the orbit center.  Some trajectories
are shown in Fig.~\ref{figlandau}. Note that they describe an \emph{ellipse}
being translated in the $x$ direction, because the confining potential is
anisotropic. Nevertheless, the motion of both projections on the ellipse axes
is harmonic, which indicates the problem will still give rise to a quantum
oscillator. Accounting for the ellipse requires a generalization of Landau's
wave function.

Inserting the solution into the Hamiltonian, the latter becomes
\begin{align}
H&=\frac{1}{2}M\omega^2R^2+\frac{1}{2}MV_x^2+\frac{1}{2}M\omega_0^2Y_0^2
\nonumber
\\
&=\frac{1}{2}M\omega^2\left(\frac{\omega^2}{\omega_c^2}X^2+Y^2\right)+
\frac{1}{2}M\omega^2\frac{\omega_0^2}{\omega_c^2}Y_0^2,
\label{hamprep}
\end{align}
where, in the second step, it has been prepared for quantization, by
collapsing the last two terms with the help of the constraint~\eqref{constr},
and expanding the radius in terms of the components along the axes.
Explicitly, the new dynamical variables are
\begin{align}
X_0&\equiv V_xt=\frac{2\omega_0^2+\omega_c^2}{2\omega^2}x-
\frac{\omega_c}{M\omega^2}p_y,
\\
Y_0&=\frac{\omega_c^2}{2\omega^2}y+\frac{\omega_c}{M\omega^2}p_x,
\\
X&\equiv x-V_xt=\frac{\omega_c^2}{2\omega^2}x+\frac{\omega_c}{M\omega^2}p_y,
\\
Y&\equiv Y_0-y=-\frac{2\omega_0^2+\omega_c^2}{2\omega^2}y+
\frac{\omega_c}{M\omega^2}p_x,
\end{align}
with the only non-zero Poisson brackets
\be
\{X,Y\}=\{X_0,Y_0\}=\frac{\omega_c}{M\omega^2}.
\ee
Standard canonical quantization in the new variables introduces
\begin{align}
\sqrt{\frac{M\omega^2}{\hbar\omega_c}}X\to \widetilde{u},\quad
\sqrt{\frac{M\omega^2}{\hbar\omega_c}}Y\to
-i\frac{\partial}{\partial\widetilde{u}}
\\
\sqrt{\frac{M\omega^2}{\hbar\omega_c}}X_0\to \widetilde{s},\quad
\sqrt{\frac{M\omega^2}{\hbar\omega_c}}Y_0\to
-i\frac{\partial}{\partial\widetilde{s}}
\end{align}
in terms of which the quantum Hamiltonian reads
\be
\widehat{H}=\frac{\hbar\omega}{2}\left(-\partial_{uu}+u^2\right)+
\frac{\hbar\omega}{2}\frac{\omega_0^2}{\omega_c^2}(-\partial_{ss}),
\ee
where the rescaling
\be
u=\sqrt{\frac{\omega}{\omega_c}}\widetilde{u},\quad
s=\sqrt{\frac{\omega}{\omega_c}}\widetilde{s},
\ee
was found convenient after quantization. The solution is evidently
\be
e^{i\nu s}\psi_n(u),
\ee
where $\nu$ is the momentum conjugate to $s$. In order to compare it with
Landau's wave function, it must be Fourier-transformed to real space. To
facilitate the comparison, all lengths are now rescaled by
$\lambda_0=\sqrt{\hbar/(M\omega)}$. Then the Fourier transform gives [cf.\
Eq.~\eqref{focksol}]
\be
\int dk_ye^{iyk_y}\exp\left[i\nu\left(
\frac{2\omega_0^2+\omega_c^2}{2\omega\omega_c}x-k_y\right)\right]
\psi_n\left(\frac{\omega_c}{2\omega}x+k_y\right)\sim
\exp\left(-i\frac{\omega_c}{2\omega}xy\right)
\exp\left(i\frac{\omega}{\omega_c}\nu x\right)
\psi_n(y-\nu)\equiv \Psi_L,
\label{lan1d}
\ee
which is indeed Landau's wave function in the limit $\omega=\omega_c$, if one
can identify $\nu=Y_0$. That identification is correct, because the
corresponding contribution to the eigenenergy
\be
E_n=\frac{\hbar\omega}{2}(2n+1)+
\frac{\hbar\omega}{2}\frac{\omega_0^2}{\omega_c^2}Y_0^2
\ee
is then identical to the classical expression~\eqref{hamprep}, given the
scaling by $\lambda_0$. This completes the derivation. One can also show
directly that the wave function~\eqref{lan1d} is the eigenfunction of the
naively quantized Hamiltonian~\eqref{ham1d}, with the same eigenvalue, as
expected.

The above derivation shows that Landau's wave function, strictly speaking, is
not a solution of the one-body problem with a 1D potential well. It is the
limiting case of such solutions, when the well potential disappears. The
classical solution physically constrains the orbit center to stop moving
($V_x=0$) when $\omega_0=0$. The quantum version of the
constraint~\eqref{constr} is the eigenvalue equation for $\partial_s$, which
reads, in real space and dimensionful variables,
\be
\widehat{V}_x\Psi_L=\frac{\omega^2}{\omega_c}Y_0\Psi_L,
\label{eigVx}
\ee
where $\widehat{V}_x$ is the Fourier transform~\eqref{lan1d} of
$(-i\hbar/M)\partial/\partial X_0$. Crucially, $\omega^2$ appears here instead
of $\omega_0^2$, so the coordinate $Y_0$ of the orbit center remains
simultaneously the wave-number for its coordinate $X_0$, when $\omega_0\to
0$.  Because of this, the wave function $\Psi_L$ retains the wave-number as a
good quantum number in that limit, albeit the associated energy vanishes as
the related translation degree of freedom becomes ``pure gauge". The
wave-number and the distance of the orbit-center coordinate $Y_0$ from the
oscillator minimum are slaved to each other, analogously to the classical
constraint~\eqref{constr}. The same operator $\widehat{V}_x$ refers to both,
even in the limit $\omega_0\to 0$.

Both Fock's and Landau's solutions for the problem in a magnetic field may
thus be viewed as limiting cases of the problem with an oscillator potential,
as that potential is turned off. They both have two quantum numbers, inherited
from the situation with non-zero potential, where the orbit-center degree of
freedom, to which the second quantum number refers, was physical. Once the
oscillator potential disappears, the two solutions differ with respect to the
charge current related to the former orbit center motion in the external
potential. In Fock's case, the wave function does not depend on the oscillator
potential at all, so it trivially retains the net clockwise orbit-center
charge current if the quantum number $n_-\neq 0$. In Landau's case, the charge
current induced by the 1D potential  vanishes with $\omega_0\to 0$, as is
known for Landau's wave function. Explicitly, for the wave functions
$\Psi_L$,
\be
\int j_x dy=-e\frac{\omega_0^2}{\omega_c}Y_0,
\ee
so that the current moves with the \emph{classical} orbit-center
velocity~\eqref{constr}. As noted above, the wave-number of the orbit center,
Eq.~\eqref{eigVx}, is not zero when $\omega_0=0$.

The above distinction underlines the independent action of the two gauge
groups in the problem. Both 2D and 1D harmonic potentials induce the breaking
of both the electromagnetic and translation gauge symmetries. When the
potential disappears, Fock's solution continues to break the electromagnetic
gauge spontaneously, while Landau's does not. Both solutions, however, still
break the translation-gauge symmetry, as evidenced by the persistence of the
second quantum number, originally related to the physical motion of the
orbit-center, induced by the external potential. In particular, the
wave-number (wavelength) of the orbit-center in Landau's wave-functions
remains (inversely) proportional to the latter's distance $Y_0$ from the
now-fictitious minimum of the vanished 1D potential. Such a spontaneous
spatial organization of Landau's basis vectors manifests the breaking of the
translation-gauge symmetry.

\section{Discussion}

It has been shown in the present work how the problem of a single electron in
a magnetic field can be quantized, and solved, without encountering the
infinite degeneracy, related to the choice of orbit center. It remains to
understand how this degeneracy crept into Landau's now-standard approach,
where one solves the naively quantized Schr\"odinger equation by direct attack
in real space. Why did Landau obtain a basis, labelled by the orbit-center
wave-number, in place of a single solution, with center fixed at a known
position, say at the origin? The analogy with the hydrogen atom in the
Introduction shows that something is indeed amiss: if one puts the proton at
the origin at the beginning, one does not expect it to delocalize by the end
of the calculation.

The classical formulations of Fock's and Landau's problems already show why
naive quantization suffices for the former, but not for the latter. Specific
linear combinations of positions and momenta, $x_0$ and $y_0$ in
Eq.~\eqref{carttrans}, have the physical meaning of orbit center coordinates.
They can thus be arbitrarily set to a constant value, say zero, in Landau's
problem, but not in Fock's, where the orbit center itself moves under
Hamiltonian action. The reason for this arbitrariness is that in Landau's
problem, the frequency $\omega_-$ in the classical Hamiltonian~\eqref{fockham}
is zero, so the orbit-center position is formally identified as a
\emph{classical} gauge degree of freedom in the sense of Dirac~\cite{Dirac50},
distinct from the electromagnetic gauge. Following Dirac's argument, because
the Poisson bracket $\{x_0,y_0\}\neq 0$, there appears a contradiction in the
naive quantization of Landau's problem, which ``blindly" promotes Poisson
brackets $\{x,p_x\}$ and $\{y,p_y\}$ to commutators with a constant value,
namely, one can construct the linear combination $[0,0]=i\hbar$.
Dirac~\cite{Dirac50} developed special methods of ``constrained quantization"
to eliminate such contradictions. In the Landau case, constrained quantization
would redefine the Poisson brackets $\{x,p_x\}$ and $\{y,p_y\}$, so as to set
their particular combination $\{x_0,y_0\}$ to zero. The present work is a
generic example of an alternative procedure, which is available whenever a
canonical transformation is known, in which the gauge variables appear as
conjugate pairs in their own right. Such is the transformation of
Eq.~\eqref{carttrans}, which isolates the translation-gauge variables $x_0$
and $y_0$. Then one can simply use standard canonical quantization in the new
variables, and observe that some conjugate pairs do not appear in the
Hamiltonian. In technical language, the gauge Hamiltonian (which is zero!) has
been uncoupled from the physical Hamiltonian. More intuitively, the problem
has been dimensionally reduced, to the subspace of only those dynamical
variables which do appear in the Hamiltonian. The contradiction $[0,0]=i\hbar$
can no longer be constructed, because $y_0$ has been replaced by
$-i\hbar\partial_{x_0}$. The only price is that non-intuitive independent
variables can appear in first quantization, such as the velocity component
$v_y$ in Eq.~\eqref{lanham}, making an integral transform necessary to obtain
the variables of choice.

Historically, Landau~\cite{Landau30} did notice that after naive quantization,
the commutator between velocity components was not zero, but used this only to
argue, correctly, that the spectrum must be that of the harmonic oscillator.
His own use of his wave functions in the calculation of the orbital magnetic
response was limited to an estimate of the one-particle density of states, for
which he had to count their degeneracy in the quantum number $k_y$ for each
oscillator state $n$, given by Eq.~\eqref{lansol}. Because each of Landau's
wave functions is exactly one Fourier component of the solution in
Eq.~\eqref{lansol}, his approach amounts to counting the number of terms in a
resolution of unity, so that his subsequent reasoning is not affected by the
present development. Remarkably, Landau's counting of the degeneracy in terms
of $x$ and $k_y$ reveals the physical insight, that these are the true
independent variables of his wave function, not $x$ and $y$, despite its
ostensibly real-space derivation. This insight is formally proven correct by
Eq.~\eqref{lansol} above.

The solution of a differential equation is physical if it satisfies the
physical symmetries and boundary conditions of the problem. In Landau's
problem, the relevant condition is that the orbit center be fixed. It is
satisfied by the coherent eigenstates of Eq.~\eqref{zzans}, as well as by the
solution~\eqref{lansol}. Notably, the latter was obtained without introducing
this condition \emph{a priori}. Canonical quantization of the one-body problem
is thus naturally free of the degeneracy associated with breaking of
translational invariance.

In the next step, when considering the many-body problem, states are
constructed by putting whole physical electrons into individual degenerate
basis states. Then a physical choice must be made, how to describe the
degeneracy. On the one side are bases which break the translation-gauge
symmetry, such as Landau's original wave functions~\cite{Landau30}, or general
($n_-\neq 0$) solutions of the Fock problem, Eq.~\eqref{focksol}. These are
delocalized in real space, as signalled by the appearance of a spurious
quantum number, linear wave-number or angular momentum of the orbit center,
respectively. On the other side is the basis of localized coherent
states~\eqref{zzans}, where the orbit center is labelled by its position, a
sharp c-number, and each basis state is by itself a physical solution of the
Landau problem. If the first choice is made, in a many-body perturbation
setting, it is to be expected on general grounds that the symmetry broken at
zeroth order is not subsequently restored. One should check in such cases,
that the spurious constants of motion in the basis vectors do not affect the
final results, or, in more technical language, that these results are properly
projected onto the physical subspace, which is invariant to both the
translation and electromagnetic gauges. Landau's and Fock's solutions are
different in this respect, because the latter contain a spurious charge
current associated with the orbit-center motion, while in the Landau case this
current vanishes.

An alternative canonical quantization of Landau's problem, shown in the
Appendix, is to decompose the motion into the coordinates of the orbit center,
as above, and the coordinates relative to the center, instead of the
velocities above. Nothing new is obtained, but it is most obvious in that
formulation, that the oscillators in Landau's and Fock's problems are simply a
quantum-physical realization of the elementary observation, that uniform
motion on a circle projects onto oscillatory motion on any axis through the
center. On the other hand, it is more immediately clear in the velocity
formulation, that the first (Landau) part of Fock's
Hamiltonian~\eqref{fockham} has no reference point in space, while the second,
in which the orbit-center coordinate $x_0$ oscillates, obviously refers to the
origin of the harmonic potential. Notably, the time-reversal symmetry breaking
inherent in the magnetic field is not manifest in the canonically quantized
Hamiltonian~\eqref{lanham}, which is real, but is innocuously hidden in the
choice, to which of the two components of velocity its variable $v$ refers:
$v_y$ ($v_x$) for a negatively (positively) charged particle.

To conclude, the problem of a single electron in a constant magnetic field has
been solved by canonical quantization in velocity space. In accord with
classical intuition, the electron cannot work radially against the effective
potential created by the magnetic field, so all its energy is in angular
motion around an arbitrary center. Its wave function has only one,
Landau-level, quantum number, physically the angular momentum around that
center, in which it is not degenerate. The degeneracy of Landau's and Fock's
solutions in a second quantum number stems from an incorrect treatment of the
arbitrariness of the center. Physically, it means that fixing the center has
not been properly separated from the calculation of the internal (one-body)
motion relative to a given center, with the consequence that a spurious second
quantum number appears, conjugate to the center position. One should not jump
to the conclusion that all work based on such wave functions was incorrect.
Much depends on how they were used, and how the result was interpreted
physically. In each particular case, it is up to the author that the
appropriate basis is selected, in light of the above results.

\acknowledgments

Conversations with S.~Bari\v{s}i\'c and E.~Tuti\v{s} are gratefully
acknowledged, as well as some pertinent comments by I.~Kup\v{c}i\'c.
This work has been supported by the Croatian Government under
Project~$0119256$.

\appendix

\section{The role of angular momentum}

\begin{figure}
\centerline{\includegraphics[height=5cm]{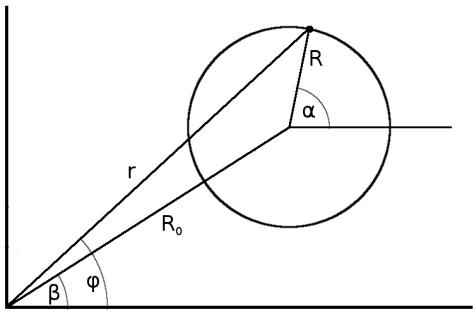}}
\caption{Classical solution of the Landau problem. Here $\alpha=\omega_ct$.
\label{figclass}}
\end{figure}
It is useful to consider the classical solution to the Landau problem in polar
coordinates in light of the above discussion. The trajectory can be written by
inspection, referring to Fig.~\ref{figclass} for the notation:
\be
r(t)=\sqrt{R_0^2+R^2+2R_0R\cos(\alpha-\beta)},\quad \alpha=\omega_ct.
\label{rsol}
\ee
Inserting this solution into the canonical equations of motion, one easily
finds the canonical angular momentum to be
\be
p_\phi=\frac{M\omega_c}{2}(R^2-R_0^2)\equiv p_\alpha-p_\beta,
\ee
where the sign of $p_\phi$ has been chosen so that the Hamiltonian gives the
known energy of rotational motion, $H=M\omega_c^2R^2/2$. These classical
results clearly show how the naive quantization $p_\phi\rightarrow
-i\partial_\phi$ admixes the orbit's fixed distance $R_0$ from an arbitrary
origin and its physical angular momentum $p_\alpha$ into the eigenvalue
$m=n_+-n_-$. A large negative $p_\phi$ simply means that one is far from the
origin (determined by the zero of the vector potential), as measured by the
spurious contribution $p_\beta$, which does not refer to anything that
moves. Conversely, $p_\phi$ does give the physical angular momentum if
$R_0=0$, when it achieves its maximum positive value. This is the classical
analogue of setting $n_-=0$ to obtain Eq.~\eqref{z00}.

After the coordinate transformation $z,\bar{z}=x\pm iy$, the quantum angular
momentum acquires the well-known form
\be
L_z=-i\partial_\phi=z\partial_z-\bar{z}\partial_{\bar{z}},
\ee
which identifies the physical angular momentum $p_\alpha$ with the
physical-motion term $z\partial_z$, while the spurious $p_\beta$ corresponds
to the gauge-motion term $\bar{z}\partial_{\bar{z}}$. But $z\partial_z$ is
just the term in the eigenvalue equation~\eqref{hzz}, or equivalently in
Eq.~\eqref{bargham}, whose eigenvalue is the quantum number $n$ in the
Landau-level spectrum. Hence all the energy of the quantum system is in the
angular momentum around the physical orbit center, corresponding to the
classical $p_\alpha$. (The analogous classical statement is
$H=\omega_cp_\alpha$.)

It is possible to quantize this problem canonically in the variables
$X=R\cos\alpha$, $Y=R\sin\alpha$, $X_0=R_0\cos\beta$, and $Y_0=R_0\sin\beta$,
in terms of which the Hamiltonian is
\be
H=\frac{M\omega_c^2}{2}\left(X^2+Y^2\right)=
\frac{\omega_c}{2}\left(\widetilde{X}^2+\widetilde{Y}^2\right),
\ee
and indeed $\{\widetilde{Y},\widetilde{X}\}=1$. Similarly
$\{X_0\sqrt{M\omega_c},Y_0\sqrt{M\omega_c}\}=
\{\widetilde{X}_0,\widetilde{Y}_0\}=1$. However, nothing new is obtained,
because there is a fixed relationship between the radius components $(X,Y)$
and the velocity components $(v_x,v_y)$, which were quantized in
Eq.~\eqref{classfock}. On the other hand, it is perhaps most obvious in this
formulation, that the harmonic oscillators in Fock's and Landau's problems are
nothing but a quantum-physical realization of the elementary observation,
that uniform motion on a circle projects to oscillatory motion on any axis
through the center.

The above quantization shows that the single half-quantum of zero-point motion
comes simply from the uncertainty of the projected position, which is
one-dimensional. If one insists on imagining the particle on the parent
circular orbit, the uncertainty smoothly oscillates between being wholly
radial near the turning-points of the projection, and wholly angular near the
mid-point. Such a separation of the uncertainty into radial and angular parts
is arbitrary, because the axis of projection may be chosen at will.

\end{document}